\documentclass{article}

\usepackage{arxiv}

\usepackage[utf8]{inputenc} 
\usepackage[T1]{fontenc}    
\usepackage{hyperref}       
\usepackage{url}            
\usepackage{booktabs}       
\usepackage{amsfonts}       
\usepackage{nicefrac}       
\usepackage{microtype}      
\usepackage{lipsum}
\usepackage{fancyhdr}       
\usepackage{graphicx}       
\graphicspath{{figures/}}     
\usepackage{CJKutf8}
\usepackage{tikz}
\usepackage{tabularx}
\usepackage{array}
\usepackage{algorithm}
\usepackage{algpseudocode} 
\usepackage{amsmath}     
\usepackage[utf8]{inputenc}
\usepackage{booktabs} 
\usepackage{tabularx} 
\usepackage{enumitem} 
\usepackage{caption}  
\usepackage{float}
\usepackage[justification=centering]{caption} 
\usepackage{subcaption} 
\usepackage{pgfplots} 
\pgfplotsset{compat=1.18} 
\usepackage[section]{placeins} 

\PassOptionsToPackage{numbers, compress}{natbib}

\makeatletter

\title{ACE-Step: A Step Towards Music Generation Foundation Model}

\author{
Junmin Gong\(^{*}\) \\
\texttt{junmin@acestudio.ai} \\
ACE Studio \\
\and
Sean Zhao \(\dagger\) \\
\texttt{sean@acestudio.ai} \\
ACE Studio \\
\and
Sen Wang \(\dagger\) \\
\texttt{sayo@acestudio.ai} \\
ACE Studio \\
\and
Shengyuan Xu \(\dagger\) \\
\texttt{shengyuan@acestudio.ai} \\
ACE Studio \\
\and
Jing Guo\(\dagger\) \\
\texttt{joe@acestudio.ai} \\
ACE Studio \\
}

\begin{document}
\maketitle
\vspace{-10.5cm} 
\begin{center}
\includegraphics[width=0.65\linewidth]{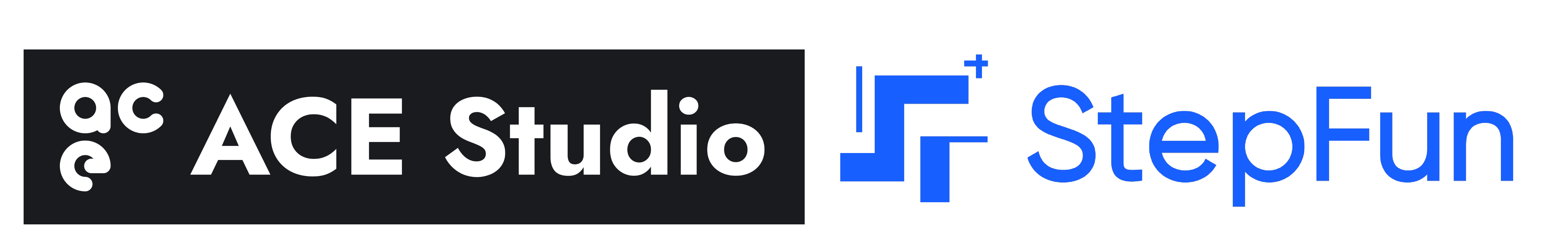}
\end{center}
\vspace{8cm} 
\begin{abstract}

We introduce ACE-Step, a novel open-source foundation model for music generation that overcomes key limitations of existing approaches and achieves state-of-the-art performance through a holistic architectural design. Current methods face inherent trade-offs between generation speed, musical coherence, and controllability. For example, LLM-based models (e.g. Yue~\cite{yuan2025yuescalingopenfoundation}, SongGen~\cite{liu2025songgen}) excel at lyric alignment but suffer from slow inference and structural artifacts. Diffusion models (e.g.DiffRhythm~\cite{ning2025diffrhythm}), on the other hand, enable faster synthesis but often lack long-range structural coherence.
ACE-Step bridges this gap by integrating diffusion-based generation with Sana’s~\cite{xie2024sana} Deep Compression AutoEncoder (DCAE~\cite{chen2024deep}) and a lightweight linear transformer. It also leverages MERT~\cite{li2023mert} and m-hubert~\cite{boito2024mhubert} to align semantic representations (REPA~\cite{yu2025repa}) during training, allowing rapid convergence. As a result, our model synthesizes up to 4 minutes of music in just 20 seconds on an A100 GPU—15× faster than LLM-based baselines—while achieving superior musical coherence and lyric alignment across melody, harmony, and rhythm metrics. Moreover, ACE-Step preserves fine-grained acoustic details, enabling advanced control mechanisms such as voice cloning, lyric editing, remixing, and track generation (e.g., lyric2vocal, singing2accompaniment).
Rather than building yet another end-to-end text-to-music pipeline, our vision is to establish a foundation model for music AI: a fast, general-purpose, efficient yet flexible architecture that makes it easy to train subtasks on top of it. This paves the way for the development of powerful tools that seamlessly integrate into the creative workflows of music artists, producers, and content creators. In short, our goal is to build a stable diffusion~\cite{rombach2021highresolution} moment for music. The code, the model weights and the demo are available at: \href{https://ace-step.github.io/}{https://ace-step.github.io/}.
\end{abstract}

\section{Introduction}
\label{sec:intro}

Artificial Intelligence Generated Content (AIGC) has achieved remarkable progress across various modalities, with models producing highly realistic text~\cite{brown2020languagemodelsfewshotlearners,touvron2023llama,qwen,deepseekai2024deepseekv3technicalreport}, images~\cite{betker2023dalle3,flux2024,imagen3_2024}, and videos~\cite{brooks2024sora_report,kling2024,wan2025}. Text-to-music generation, however, presents unique challenges due to the complex interplay of melody, harmony, rhythm, timbre, long-range structure, and linguistic content in the form of lyrics.

Early work like OpenAI's Jukebox~\cite{dhariwal2020jukebox} demonstrated the feasibility of generating raw audio music with vocals, albeit with significant computational cost and coherence limitations. Subsequent research explored various architectures, including Transformer-based models like MusicGen~\cite{copet2024simplecontrollablemusicgeneration} using quantized audio representations, and diffusion models like AudioLDM~\cite{liu2023audioldm,audioldm2-2024taslp} operating in latent spaces for high-fidelity synthesis.

Recently, commercial systems like Suno~\cite{suno2024}, Udio~\cite{udio2024}, and Riffusion~\cite{riffusion2022} have significantly advanced the state-of-the-art. Suno, in particular, has garnered attention for generating full songs with strong coherence, style alignment, musicality, and lyric adherence, effectively reshaping music creation by integrating composition and production into streamlined platforms. While this democratizes music creation, it also raises concerns about traditional industry workflows and artists' livelihoods.

In response, the open-source community has produced models such as Yue~\cite{yuan2025yuescalingopenfoundation}, SongGen~\cite{liu2025songgen}, and DiffRhythm~\cite{ning2025diffrhythm}. However, these often fall short of commercial counterparts, suffering from limitations in musical quality, poor style generalization, weak lyric-audio alignment, slow inference speeds, or reliance on specific input formats that hinder practical usability. Unlike text, image, and video domains where open-source models achieve competitive performance, the text-to-music landscape lags considerably behind.

This disparity stems from several factors: scarcity of large-scale, high-quality annotated music datasets with precise lyric alignment, the inherent difficulty of simultaneously modeling musical structure and vocal characteristics, and lack of comprehensive evaluation benchmarks. Current approaches follow two main paradigms: (1) two-stage methods~\cite{dhariwal2020jukebox} that generate discrete codes before audio synthesis, suffering from error propagation and slow sequential generation; and (2) diffusion models~\cite{evans2024stableaudioopen,novack2025fasttexttoaudiogenerationadversarial,ning2025diffrhythm} operating on compressed latent representations, offering higher fidelity but requiring specific techniques for structural coherence and controllability.

To bridge this critical gap, we introduce ACE-Step, a novel foundation model for text-to-music synthesis. Our work aims to provide a powerful, efficient, and versatile open-source alternative that empowers creators rather than replacing them. We believe continuous open-source iteration is crucial for creating accessible tools that augment human creativity and democratize musical creation. Drawing inspiration from recent advancements in diffusion models, particularly in text-to-image domain, ACE-Step introduces several key innovations:

\begin{enumerate}
\item \textbf{Novel Architectural Integration:} We leverage Deep Compression AutoEncoder (DCAE~\cite{chen2024deep}) to achieve highly compact mel-spectrogram latent representation, synergistically integrated with flow matching~\cite{liu2022flow} generative process driven by a linear transformer backbone~\cite{cai2024efficientvitmultiscalelinearattention}. This combination significantly accelerates both training and inference for generating coherent, multi-minute musical pieces.

\item \textbf{Enhanced Semantic Alignment:} We incorporate Representation Alignment (REPA~\cite{yu2025repa}) technique by applying pre-trained MERT~\cite{li2023mert} (for music representation) and mHuBERT~\cite{boito2024mhubert} (for multilingual speech representation) as semantic guides during flow matching training, promoting faster convergence and improved semantic fidelity, especially in stylistic accuracy and lyric adherence.

\item \textbf{Advanced Control via Flow Manipulation:} Inspired by controllable image synthesis, ACE-Step implements fine-grained creative control including musical variations, audio repainting (temporal inpainting), and precise lyric editing through flow-based manipulation principles.

\item \textbf{Foundation Model Extensibility:} We demonstrate ACE-Step's versatility through successful downstream applications, including LoRA~\cite{hu2021loralowrankadaptationlarge} and ControlNet-inspired~\cite{zhang2023addingconditionalcontroltexttoimage} conditioning for tasks such as direct lyric-to-vocal synthesis and targeted text-to-sample generation.

\item \textbf{Practical Robustness:} ACE-Step is engineered for real-world application with user-specifiable output duration and robustness to diverse input formats for both descriptive tags and lyrics.
\end{enumerate}

\section{Related Work}
\label{sec:related_work}

This section reviews prior work in audio and music generation, highlighting key open-source and closed-source models, and detailing the foundational generative techniques and components leveraged by ACE-Step.

\subsection{Generative Modeling for Audio and Music}
Generative audio modeling, encompassing speech and environmental sounds, is a well-established field. Text-to-Speech (TTS) synthesis, for example, has evolved from concatenative/parametric methods~\cite{black2007statisticalparametricspeechsynthesis} to neural network-based systems, often employing two-stage pipelines (acoustic model to intermediate representation, then vocoder to waveform) or, more recently, diffusion models~\cite{jiang2025sparse, liu2023audioldm, audioldm2-2024taslp} for high-fidelity output.

Music generation introduces distinct challenges: complex harmony, rhythm, long-range structure, timbre, and lyric alignment. Two dominant paradigms have emerged:
\begin{enumerate}
    \item \textbf{Two-Stage Discrete Tokenization:} Pioneered by Jukebox~\cite{dhariwal2020jukebox}, this involves an autoencoder (e.g., VQ-VAE~\cite{oord2018neuraldiscreterepresentationlearning}) compressing audio into discrete tokens, which an auto-regressive model (often a Transformer) then predicts sequentially. This can model long dependencies but may suffer from tokenization information loss and slow sequential inference.
    \item \textbf{Latent Diffusion Models:} Inspired by image generation, these models apply diffusion processes to a compressed latent audio representation (typically mel-spectrograms) learned by an autoencoder~\cite{agostinelli2023musiclmgeneratingmusictext, copet2024simplecontrollablemusicgeneration}. Examples like Stable Audio~\cite{evans2024stableaudioopen} and Riffusion~\cite{riffusion2022} often yield high-fidelity audio and faster synthesis, though ensuring long-range coherence and controllability can require specific designs.
\end{enumerate}

\subsection{Open-Source Works}
Several open-source models have advanced text-to-music generation:
\textbf{Jukebox}~\cite{dhariwal2020jukebox}, a pioneering model, generates raw audio with vocals using hierarchical VQ-VAEs and autoregressive Transformers, offering diversity but suffering from slow inference.
\textbf{MusicGen}~\cite{copet2024simplecontrollablemusicgeneration} employs a single-stage autoregressive Transformer on EnCodec~\cite{défossez2022highfidelityneuralaudio} tokens, improving speed and enabling melody conditioning and lyric integration.
\textbf{AudioLDM}~\cite{liu2023audioldm, audioldm2-2024taslp} pioneered latent diffusion for general audio, including music, using CLAP~\cite{laionclap2023} for text conditioning, with AudioLDM 2 enhancing musical structure and lyric coherence.
\textbf{Stable Audio Open}~\cite{evans2024stableaudioopen} utilizes a diffusion model over autoencoded latent codes, conditioned by T5~\cite{2020t5}, to generate longer, high-quality audio, emphasizing practical usability.
\textbf{Yue}~\cite{yuan2025yuescalingopenfoundation} is an open-source foundation model aiming for full-length song generation with coherent lyrics and melody.
\textbf{SongGen}~\cite{liu2025songgen} uses a single-stage autoregressive Transformer for text-to-song generation, focusing on vocal-instrumental harmony.
\textbf{DiffRhythm}~\cite{ning2025diffrhythm} introduces a latent diffusion framework for efficient full-song synthesis, with a focus on rhythm-aware modeling to ensure temporal coherence and nuanced expressive control.

\subsection{Closed-Source Works and Our Motivation}
Closed-source text-to-music generation has also advanced rapidly. Early efforts like Riffusion~\cite{riffusion2022} adapted image diffusion models (e.g., Stable Diffusion~\cite{rombach2021highresolution}) to mel-spectrograms, showing potential but facing fidelity challenges. More recent systems often employ sophisticated multi-stage architectures. Google's MusicLM~\cite{agostinelli2023musiclmgeneratingmusictext} uses hierarchical sequence modeling for high-fidelity, long-form audio. Models like Suno~\cite{suno2024} and Udio~\cite{udio2024} are understood to use LLM-based hierarchical processes, first generating semantic tokens then refining them into audio, offering strong coherence and control, though their proprietary nature limits detailed comparison and community experimentation.

This review highlights a dynamic landscape: foundational autoregressive models (Jukebox), faster variants (MusicGen, SongGen), and the rise of diffusion models (AudioLDM, Stable Audio) for fidelity. While closed-source systems like Suno produce impressive "one-shot" complete songs, this may limit creative intervention.
ACE-Step, conversely, adopts a diffusion-based architecture, inspired by successes in image synthesis. This choice aims to foster a collaborative creation ecosystem, enabling fine-grained control (cf. ControlNet~\cite{zhang2023addingconditionalcontroltexttoimage}), customization (cf. LoRA~\cite{hu2021loralowrankadaptationlarge}), and iterative refinement. Our goal is to develop a "Stable Diffusion for Music"—a powerful, flexible, and open foundation that empowers creators.

\subsection{Core Generative Techniques and Components}
\label{subsec:core_techniques}
Our work builds upon several advanced generative modeling paradigms and representation learning techniques. Understanding their principles is key to appreciating their application within ACE-Step.

\begin{enumerate}
    \item \textbf{Flow Matching (FM):}
    Flow matching~\cite{liu2022flow} offers a simulation-free approach to training continuous normalizing flows (CNFs). Unlike score-based diffusion models that learn the gradient of the log-density (score function) $\nabla \log p_t(x)$, FM directly regresses a time-dependent vector field $v_t(x)$ that transports samples from a simple prior distribution $p_0$ to a complex target data distribution $p_1$ (or vice-versa) along probability paths $p_t$. This is typically achieved by minimizing a loss between the model's predicted vector field and a target vector field defined by samples from $p_0$ and $p_1$. This direct regression can lead to simpler training objectives, potentially faster convergence, and more stable training dynamics compared to score matching. Conditional Flow Matching (CFM)~\cite{lipman2024flowmatchingguidecode} extends this by allowing the vector field to depend on conditioning information, enabling guided generation. Sampling involves numerically solving an ordinary differential equation (ODE) defined by the learned vector field.

    \item \textbf{Diffusion Transformers (DiT):}
    The Diffusion Transformer (DiT)~\cite{peebles2023scalablediffusionmodelstransformers} architecture adapts the Transformer~\cite{vaswani2023attentionneed} to serve as the backbone for diffusion models, primarily replacing U-Net~\cite{ronneberger2015unetconvolutionalnetworksbiomedical} structures common in image generation. DiTs treat latent representations (e.g., image patches or, in our domain, segments of audio features) as sequences of tokens. These tokens, along with embeddings for the diffusion timestep and any conditioning information, are processed by a series of Transformer blocks. The Transformer's inherent strengths in modeling long-range dependencies, its attention mechanism's ability to capture global context, and its proven scalability make it a powerful choice for diffusion backbones, especially as model and dataset sizes increase. Linear attention variants~\cite{cai2024efficientvitmultiscalelinearattention} can further enhance efficiency for very long sequences.

    \item \textbf{Representation Alignment (REPA):}
    Representation Alignment~\cite{yu2025repa} encompasses techniques designed to bridge semantic gaps between representations learned by different models or derived from different modalities. The core idea is to transform or guide one representation space to be more semantically consistent with another, often a powerful pre-trained "teacher" model's embedding space. This alignment is typically enforced during training by adding a loss term that encourages similarity (e.g., via cosine similarity or MSE) between the student model's intermediate or output representations and the target representations from the teacher model for corresponding inputs. REPA can facilitate knowledge transfer, improve conditional generation by enforcing better adherence to semantic guidance, and potentially accelerate training by providing richer supervisory signals.

    \item \textbf{Deep Compression AutoEncoders (DCAE) for Audio:}
    Efficiently modeling complex, high-dimensional data like audio often necessitates an initial dimensionality reduction step. Autoencoders, particularly those designed for significant compression like the Deep Compression AutoEncoder (DCAE)~\cite{chen2024deep}, are crucial for this. A DCAE learns an encoder to map high-dimensional input (e.g., mel-spectrograms) to a much lower-dimensional latent representation, and a decoder to reconstruct the original input from this latent code. The "deep compression" aspect implies a focus on achieving a highly compact latent space while minimizing reconstruction error. For audio, this means capturing salient acoustic features essential for perception and quality within a small number of latent variables. This not only reduces the computational burden for subsequent generative models operating in this latent space but also encourages the generative model to focus on higher-level structural and semantic aspects rather than low-level waveform details. The specific architecture of DCAE (e.g., convolutional layers, quantization if used) is optimized for this trade-off.
\end{enumerate}
\section{Method}
\label{sec:method}

ACE-Step is designed as a fast, general-purpose, and flexible foundation model for music generation. We adopt a diffusion-based generation paradigm, integrating efficient architectural choices and advanced semantic alignment techniques. This section details the data foundation, model architecture, and training methodology.

\subsection{Data Foundation}
The performance of ACE-Step is built upon a large-scale, diverse audio dataset, meticulously curated, processed, and annotated to provide rich conditioning signals.

\subsubsection{Dataset Curation and Quality Control}
Our primary training corpus comprises approximately 1.8 million unique musical pieces (roughly 100,000 hours), spanning 19 languages, with a majority in English. To ensure data quality, we employed the Audiobox aesthetics toolkit~\cite{tjandra2025aes}, a suite of audio processing tools, for quality assessment and filtering. This process identified and removed low-fidelity recordings and live performances, which often contain undesirable artifacts.

\subsubsection{Annotation and Conditioning Signals}
Comprehensive annotations were generated to provide rich conditioning signals for the model:
\begin{itemize}
    \item \textbf{Descriptive Captions:} The Qwen-omini model~\cite{Qwen2.5-Omni} generated textual descriptions detailing musical content and mood.
    \item \textbf{Lyrics Transcription and Alignment:} Vocals were transcribed using Whisper 3.0~\cite{radford2022robustspeechrecognitionlargescale}. A locality-sensitive hashing (LSH)~\cite{ekzhu_datasketch_2024} based approach was then used to map the International Phonetic Alphabet (IPA)~\cite{zhu2022charsiu-g2p} representations of these transcriptions to canonical lyric databases for improved alignment. 
    \item \textbf{Structural Segmentation:} High-level song structures (e.g., intro, verse, chorus) were identified using an "all-in-one" music understanding model~\cite{taejun2023allinone}.
    \item \textbf{Musical Attributes:} Beats Per Minute (BPM) were extracted using `Beatthis`~\cite{foscarin2024beatthis}, while `Essentia`~\cite{essentia2013} provided key and initial stylistic tags. These tags were subsequently refined and "recaptioned" into natural language phrases by Qwen-omini for consistency.
\end{itemize}

\subsubsection{Input Strategies and Robustness}
To enhance model robustness and adaptability to diverse inputs, we implemented several key strategies:
\begin{itemize}
    \item \textbf{Conditional Lyric Handling:} For tracks with vocals but missing lyrics, empty strings were used as input. Purely instrumental tracks were explicitly marked with special tokens like `[inst]` or `[instrumental]`. Lyric tokenization follows the methodology of SongGen~\cite{liu2025songgen}, utilizing the XTTS VoiceBPE tokenizer~\cite{casanova2024xttsmassivelymultilingualzeroshot} for its broad multilingual support. While Romanized languages undergo minimal preprocessing, non-Romanized scripts (e.g., Chinese, Japanese, Korean) are converted to phonemic representations using grapheme-to-phoneme (G2P) tools prior to tokenization. This approach facilitates fine-grained editing (e.g., specifying Pinyin for Chinese polyphonic characters) while maintaining tokenization consistency.
    \item \textbf{Variable-Length Training:} To support user-specified output durations, ACE-Step was trained on variable-length audio segments. A custom dataset sampler grouped segments of similar lengths into batches, mitigating training inefficiencies arising from length variability.
    \item \textbf{Diverse Textual Prompting:} The model was trained with diverse textual annotations, including comma-separated tags, descriptive captions, and usage scenarios, which were dynamically generated by an audio language model to match context. To further prevent overfitting to specific prompt formats, lyrics were augmented by an LLM into multiple stylistic variations (e.g., with or without structured tags), enhancing adaptability to real-world input inconsistencies.
\end{itemize}

\subsubsection{Addressing Lyric-Audio Alignment Challenges}
Achieving accurate full-song lyric alignment presented significant challenges due to musical rhythms, pauses, and annotation inaccuracies (e.g., uncredited backing vocals, transcription errors). Our approach evolved iteratively:
\begin{itemize}
    \item \textbf{Initial Attempts and Limitations:} Early experiments using the International Phonetic Alphabet (IPA) for direct lyric representation proved impractical for user editing and cross-lingual phoneme management. Furthermore, attempting to directly link lyric semantics to low-level pronunciation characteristics introduced noise and hindered learning efficiency, suggesting an insufficient correlation for robust alignment without stronger guidance.
    \item \textbf{The Role of Semantic Guidance (REPA):} The Representation Alignment (REPA) framework (detailed in Section~\ref{subsec:core_techniques} became crucial. Diffusion models tend to prioritize learning low-frequency structures (e.g., melody, song form) before high-frequency details (e.g., precise vocal nuances). Without strong semantic guidance from pre-trained Self-Supervised Learning (SSL) models (mHuBERT~\cite{boito2024mhubert} for speech, MERT~\cite{li2023mert} for music), the model could prematurely focus on vocal details leading to misaligned or omitted lyrics. REPA enforces sustained semantic constraints from these SSL models, ensuring lyric adherence is treated as a fundamental, low-frequency priority throughout the generation process.
    \item \textbf{Validation:} Ablation studies confirmed REPA’s necessity: models trained without REPA exhibited severe alignment degradation. Even when alignment showed some improvement with extended training, removing REPA during fine-tuning caused a notable decay in performance, underscoring its critical role in maintaining robust lyric delivery.
\end{itemize}

\subsection{Model Architecture}
\label{subsec:model_architecture}
\begin{figure}[H] 
  \centering
  \includegraphics[width=1\linewidth]{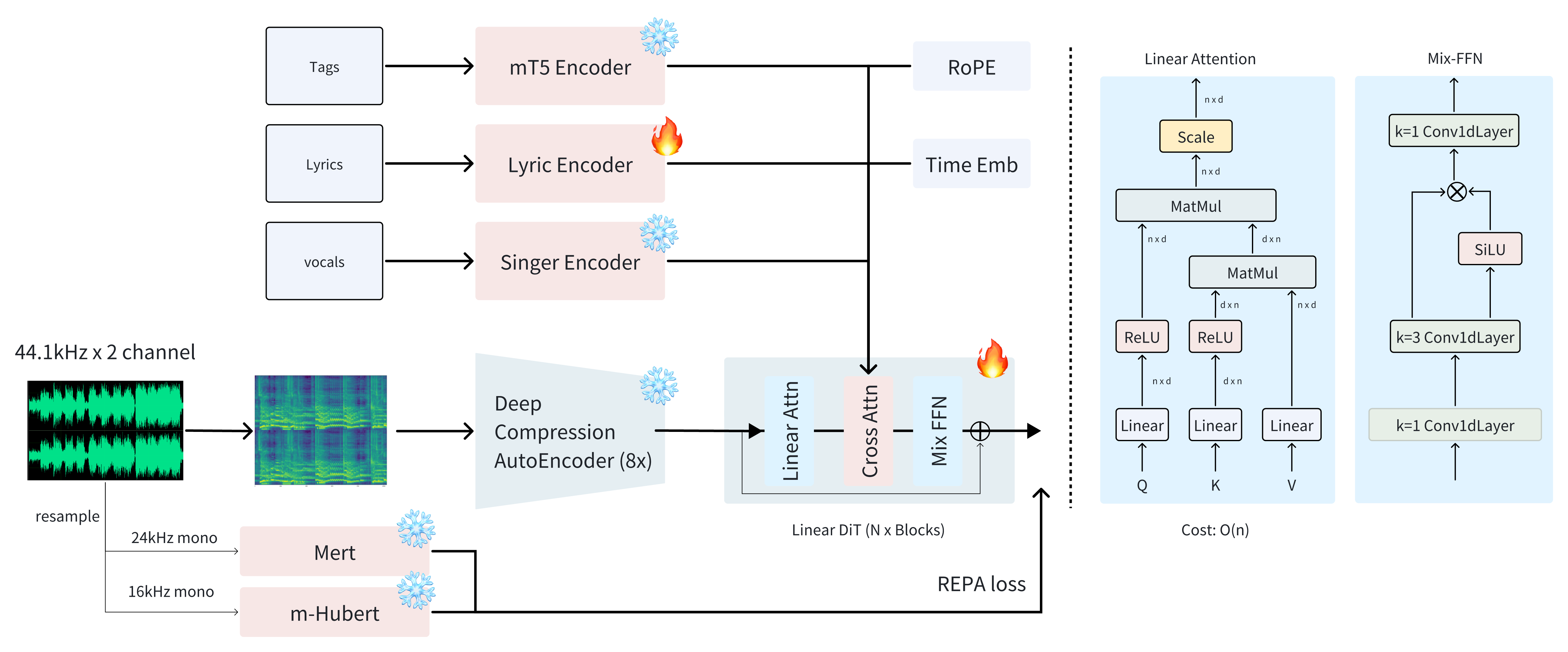}
  \caption{ACE-Step Framework}
  \label{fig:ace-step}
\end{figure}

As shown in Figure~\ref{fig:ace-step}, ACE-Step adapts a text-to-image diffusion framework, drawing inspiration from architectures like Sana~\cite{xie2024sana}, for music generation. The core generative model is a diffusion model operating on a compressed mel-spectrogram latent representation. This process is guided by conditioning information from three specialized encoders: a text prompt encoder, a lyric encoder, and a speaker encoder. Embeddings from these encoders are concatenated and integrated into the diffusion model via cross-attention mechanisms.

\subsubsection{Audio Autoencoder: Music-DCAE}
For efficient latent space modeling, we employ a Deep Compression AutoEncoder (DCAE). The architecture largely follows previous work ~\cite{chen2024deep}, but with a critical modification for mel-spectrogram compression. An initial configuration using 32x compression in both time and frequency dimensions (f32c32, channel=32) resulted in unacceptable audio quality degradation. We subsequently adopted an 8x compression setting (f8c8, channel=8, yielding approximately 10.77Hz temporal resolution in the latent space), which provided a superior balance between compression ratio and fidelity. For converting the generated mel-spectrograms back to waveform (vocoding), we utilize a pre-trained universal music vocoder from Fish Audio~\cite{fish-speech-v1.4}.

We also conducted preliminary explorations with 1D VAE architectures, including variants similar to those in Stable Audio~\cite{evans2024stableaudioopen} and DiffRhythm~\cite{ning2025diffrhythm}. However, these approaches, possibly due to suboptimal integration with our diffusion model's training recipe, exhibited persistent high-frequency artifacts in the generated vocals, particularly in later training stages. Given project timelines, these alternatives would be pursued further in the next version.

\subsubsection{Core Diffusion Model: Linear DiT}
The denoising network at the heart of ACE-Step is a Linear Diffusion Transformer (DiT). We adapt the Linear DiT structure from Sana~\cite{xie2024sana} with two key modifications:
\begin{itemize}
    \item \textbf{Simplified Adaptive Layer Normalization:} We replace the standard AdaLN with AdaLN-single~\cite{chen2023pixartalphafasttrainingdiffusion}, where a single AdaLN layer's parameters are shared across all DiT blocks. This significantly reduces model size and memory consumption.
    \item \textbf{1D Convolutional FeedForward Layers:} The FeedForward Network (FFN) layers within the DiT blocks are modified from using 2D convolutions to 1D convolutions. This change better aligns the architecture with the sequential, 1D nature of the temporal information in the audio latent space after patchification.
\end{itemize}

\subsubsection{Conditioning Encoders}
To guide the generation process, the Linear DiT is conditioned on embeddings from the following encoders:
\begin{itemize}
    \item \textbf{Text Encoder:} We employ a frozen Google mT5-base model~\cite{chung2023unimaxfairereffectivelanguage} to generate 768-dimensional embeddings from textual prompts. This model was chosen for its robust multilingual capabilities in processing both descriptive tags and natural language captions.
    \item \textbf{Lyric Encoder:} The lyric encoder architecture and hyperparameters are adopted from SongGen~\cite{liu2025songgen}. This encoder is trainable during ACE-Step's training process.
    \item \textbf{Speaker Encoder:} The speaker encoder processes a 10 - second unaccompanied vocal segment, which is separated by demucs, into a 512 - dimensional embedding. For full songs with vocals, embeddings from multiple such segments are averaged. A zero vector is used as the speaker embedding for instrumental tracks. The encoder, pre-trained on a large and diverse singing voice corpus, draws architectural inspiration from PLR-OSNet~\cite{xie2020learningdiversefeaturespartlevel}, originally designed for face recognition. During the main training phase of ACE-Step, a 50\% dropout rate is applied to these speaker embeddings. This prevents the model from over-relying on timbre information for stylistic interpretation, thereby enabling reasonable timbre generation even without explicit speaker input. In a subsequent fine-tuning stage, speaker embeddings are omitted entirely. While this degrades precise voice cloning capabilities, it transfers greater stylistic control to the text prompts, enhancing overall flexibility.
\end{itemize}

\subsection{Training Objectives}
\label{subsec:training_objectives}
ACE-Step is trained by optimizing a composite objective function that balances generative fidelity with semantic coherence. This involves a primary flow matching loss for learning the data distribution in the latent space, and an auxiliary semantic alignment loss leveraging pre-trained Self-Supervised Learning (SSL) models. Our overall training strategy, including learning rate schedules and optimizers, draws from established practices in training large-scale generative models, such as those for Stable Diffusion 3~\cite{esser2024scalingrectifiedflowtransformers}.

\subsubsection{Flow Matching Loss}
The core generative capability of ACE-Step is learned through a continuous-time flow matching (FM) objective~\cite{lipman2024flowmatchingguidecode}.
Given a clean audio segment, it is first converted to a mel-spectrogram and then encoded by the pre-trained Music-DCAE (Section~\ref{subsec:model_architecture}) into a latent representation $x_0$.
We define a linear probability path that transports samples from a simple Gaussian noise distribution $z \sim \mathcal{N}(0,I)$ to the target data distribution $x_0$.
During training, a time $t \sim U[0,1]$ is sampled. A coefficient $\sigma_t$ is derived from this $t$.
A noisy latent $x_{\text{noisy}}$ is constructed via interpolation: $x_{\text{noisy}} = (1-\sigma_t) x_0 + \sigma_t z$.
The model $\epsilon_\theta$, conditioned on $x_{\text{noisy}}$, the sampled time $t$, and various embeddings (text, speaker, lyric, denoted collectively as $\text{condition}$), is trained to predict a vector $v_{\text{out}} = \epsilon_\theta(x_{\text{noisy}}, t, \text{condition})$.
The target for this predicted vector $v_{\text{out}}$ is $-(x_0-z)$. This target can be interpreted as the negative of the constant velocity field $v = x_0-z$ associated with the linear path from $z$ to $x_0$.
The clean latent $x_0$ is then estimated from $x_{\text{noisy}}$ and $v_{\text{out}}$ using the preconditioning formula: $x_0^{\text{pred}} = v_{\text{out}} \cdot (-\sigma_t) + x_{\text{noisy}}$.
The flow matching loss $\mathcal{L}_{\text{FM}}$ is then the Mean Squared Error (MSE) between this reconstructed prediction $x_0^{\text{pred}}$ and the ground truth $x_0$:
$$ \mathcal{L}_{\text{FM}} = \mathbb{E}_{x_0, z, t} [\| (\epsilon_\theta(x_{\text{noisy}}, t, \text{condition}) \cdot (-\sigma_t) + x_{\text{noisy}}) - x_0 \|_2^2 ] $$
where $x_{\text{noisy}} = (1-\sigma_t) x_0 + \sigma_t z$.

\subsubsection{Semantic Alignment Loss}
To enhance musical coherence, particularly the alignment and intelligibility of sung lyrics, we incorporate an auxiliary semantic alignment loss, inspired by the Representation Alignment (REPA) framework~\cite{yu2025repa}. This loss constrains intermediate representations from our model's Linear DiT backbone to align with those from powerful pre-trained audio SSL models.

Specifically, we extract features $h_{\text{DiT}}$ from an intermediate layer of ACE-Step's Linear DiT (the 8th layer out of 24). These features are then aligned with target representations obtained from:
\begin{itemize}
    \item \textbf{MERT (Music Encoder Representations from Transformers)~\cite{li2023mert}:} This model provides general musical feature embeddings. We process the original audio (resampled to 24kHz, mono) in 5-second chunks. MERT outputs embeddings $h_{\text{MERT}}$ of dimension $1024 \times T_M$, where $T_M$ is the number of temporal frames at a rate of 75Hz.
    \item \textbf{mHuBERT (multilingual HuBERT)~\cite{boito2024mhubert}:} This model yields representations beneficial for lyric intelligibility and speech-like characteristics. The original audio (resampled to 16kHz, mono) is processed in 30-second chunks. mHuBERT outputs embeddings $h_{\text{mHuBERT}}$ of dimension $768 \times T_H$, where $T_H$ is the number of temporal frames at a rate of 50Hz.
\end{itemize}

Before computing the alignment loss, the temporal dimensions of $h_{\text{DiT}}$, $h_{\text{MERT}}$, and $h_{\text{mHuBERT}}$ are matched, typically through interpolation or pooling, to a common length $T'$. The SSL loss is the average of the 1 - cosine distance between the DiT's representations and the SSL models' representations:
$$ \mathcal{L}_{\text{SSL}} = \frac{1}{2} \left( \text{cosineSim}(h'_{\text{DiT}}, h'_{\text{MERT}}) + \text{cosineSim}(h'_{\text{DiT}}, h'_{\text{mHuBERT}}) \right) $$
where $h'$ denotes the temporally aligned representations. For instance, $h'_{\text{DiT}}$ represents the DiT features after temporal alignment.

The final training objective is a weighted sum of the flow matching and semantic alignment losses:
$$ \mathcal{L}_{\text{Total}} = \mathcal{L}_{\text{FM}} + \lambda_{\text{SSL}} \cdot \mathcal{L}_{\text{SSL}} $$
where $\lambda_{\text{SSL}}$ is a hyperparameter controlling the influence of the SSL guidance, empirically set to 1.0 in our experiments.

\section{Experiments}
\label{sec:experiments}

We trained the Music-DCAE and ACE-Step models using DeepSpeed ZeRO Stage 2. During pre-training, we set the SSL loss weight $\lambda_{\text{SSL}}$ to 1.0, but reduced the mHuBERT SSL loss component to 0.01 for the final 100,000 fine-tuning steps to prevent suppression of instrumental sounds and preserve musicality. Due to the lack of an automated evaluation pipeline, we relied on manual listening assessments every 2,000 steps to monitor performance, which limited our ability to precisely track training progress or detect overfitting. Detailed training configurations are provided in Tables~\ref{tab:dcae_training_details} and~\ref{tab:ace_step_training_details} in Appendix~\ref{sec:appendix_training_details}.



\section{Evaluation}
\label{sec:evaluation}

\subsection{Human Evaluation}
\label{ssec:human_eval}

\begin{figure}[H]
  \centering
  \includegraphics[width=0.65\linewidth]{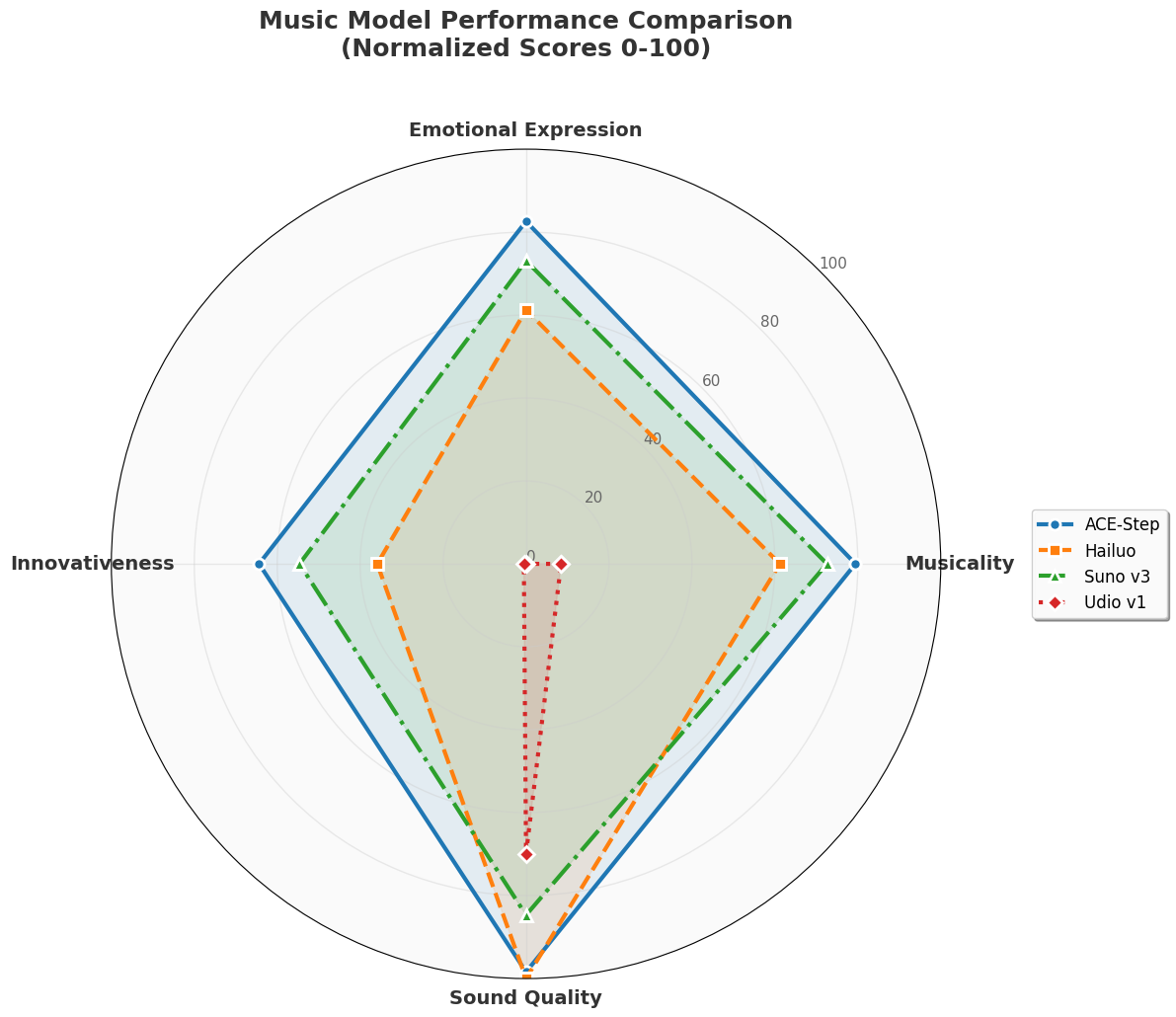}
  \caption{Human evaluation results comparing ACE-Step against three baseline models (Hailuo, Suno v3, and Udio v1) across four dimensions. Scores are normalized to 0-100.}
  \label{fig:human_eval}
\end{figure}

We conducted a blind human evaluation with 32 participants to assess ACE-Step's performance. The cohort comprised 7 music professionals, 7 enthusiasts, and 18 general listeners (15 female, 17 male), aged 19-42 years. Participants evaluated generated music samples using a 5-point Likert scale across four dimensions:

\begin{itemize}
    \item \textbf{Musicality:} Melody/harmony logic, rhythmic complexity, and structural integrity
    \item \textbf{Emotional Expression:} Clarity of emotion, resonance, and dynamic contrast
    \item \textbf{Innovativeness:} Originality in style, timbre, and structural elements
    \item \textbf{Sound Quality:} Audio fidelity, spatial characteristics, and absence of artifacts
\end{itemize}

ACE-Step was compared against three closed-source baselines: Suno v3, Udio v1, and Hailuo. All evaluations used randomized sample presentation to mitigate order effects. As shown in Figure~\ref{fig:human_eval}, ACE-Step achieves strong performance with scores of approximately 85 in Emotional Expression, 82 in Innovativeness, 80 in Sound Quality, and 78 in Musicality. Suno v3 follows with balanced scores around 70-75 across all dimensions, while Hailuo shows slightly lower but comparable performance (65-72). Udio v1 significantly underperforms with scores below 15 in all categories.

While this evaluation excluded newer models (Suno v3.5/v4, Tiangong, Mureka), community feedback suggests ACE-Step's performance falls between Suno v3 and v3.5, remaining competitive among open-source alternatives despite being surpassed by the latest commercial offerings.

\begin{table}[H]
\centering
\caption{Waveform Reconstruction Performance. FAD scores are lower is better. Audiobox-aesthetics scores are higher is better.}
\label{tab:waveform_recon}
\begin{tabular}{l|ccc|cccc}
\toprule
Model & \multicolumn{3}{c|}{FAD$\downarrow$} & \multicolumn{4}{c}{Audiobox-aesthetics$\uparrow$} \\
 & cdpam-acoustic & cdpam-content & dac-44kHz & CE & CU & PC & PQ \\
\midrule
Music DCAE (mel 2d) & 0.0224 & 0.0202 & 49.3879 & 6.5511 & 7.2299 & 6.0457 & 7.4611 \\
DiffRhythm VAE (1d) & 0.0059 & 0.0048 & 22.3142 & 6.7164 & 7.3759 & 5.9301 & 7.6305 \\
\bottomrule
\end{tabular}
\end{table}

\label{sssec:song_gen_metrics}
\begin{figure}[H]
  \centering
  \begin{subfigure}[b]{0.49\textwidth}
    \centering
    \includegraphics[width=\linewidth]{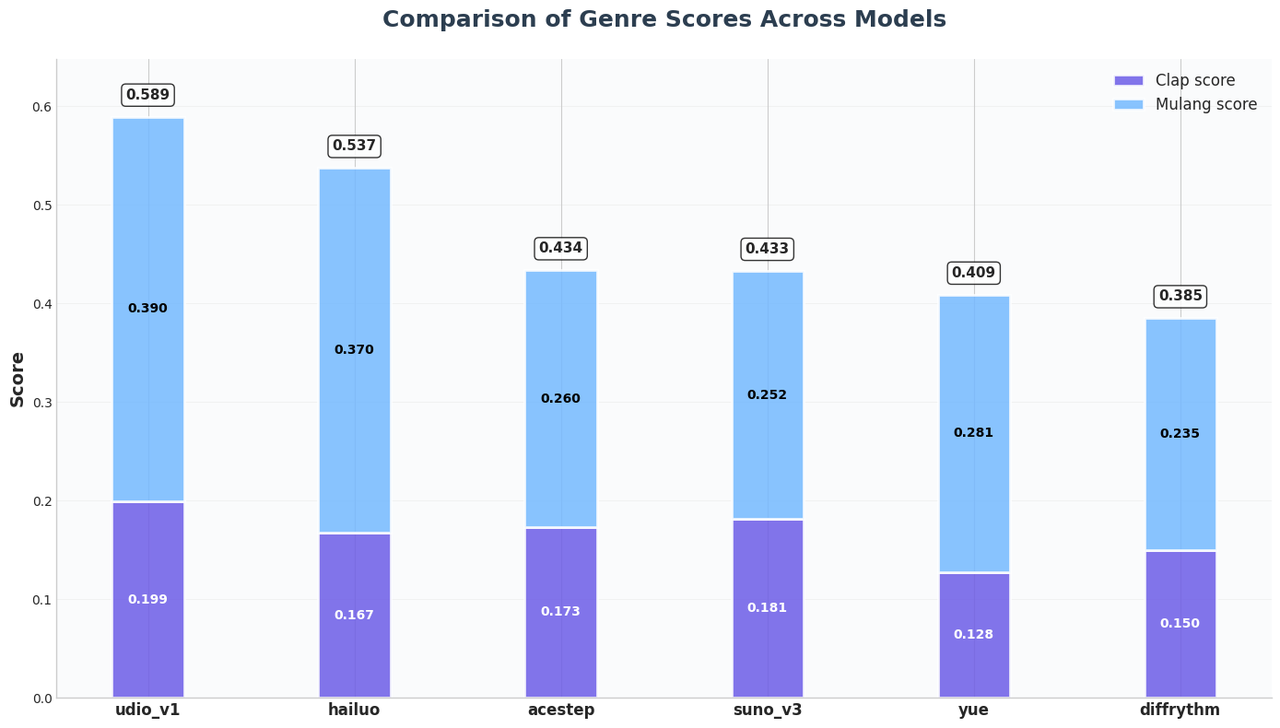} 
    \caption{Genre Scores (CLAP + Mulan). Higher is better.}
    \label{fig:genre_sub_score}
  \end{subfigure}
  \hfill 
  \begin{subfigure}[b]{0.49\textwidth}
    \centering
    \includegraphics[width=\linewidth]{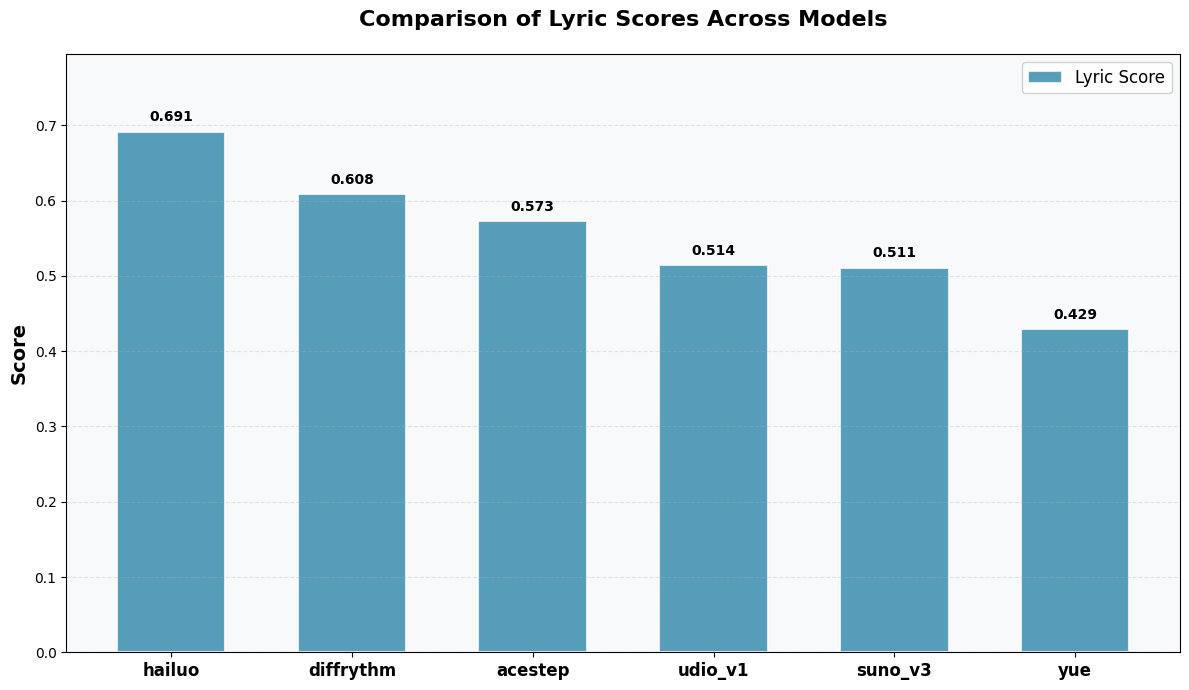} 
    \caption{Lyric Scores (Whisper Forced Alignment). Higher is better.}
    \label{fig:lyric_sub_score}
  \end{subfigure}
  \caption{Comparison of (a) Genre Scores and (b) Lyric Scores Across Models.}
  \label{fig:style_lyric_scores}
\end{figure}

\subsection{Automatic Evaluation}
\label{ssec:auto_eval}
Automatic evaluations were conducted across five main aspects: waveform reconstruction, style alignment, lyric alignment, aesthetic quality (Audiobox), and musicality (SongEval), as well as generation speed. For song generation tasks, we utilized a set of 20 prompts ~\ref{sec:eval_prompts_appendix} (10 in English, 10 in Chinese) covering diverse musical styles, generated by an LLM, which included style tags and lyrics.

\subsubsection{Waveform Reconstruction}
\label{sssec:waveform_recon}

To evaluate waveform reconstruction, we randomly selected 20 ground truth songs. These songs were then reconstructed using both the Music DCAE and DiffRhythm's VAE. We subsequently calculated the Fr\'echet Audio Distance (FAD) and aesthetic scores between the reconstructions and their respective ground truth originals. For FAD, we utilized the \texttt{fadtk} package~\cite{fadtk}. Aesthetic quality was assessed using Meta's Audiobox-aesthetics framework~\cite{tjandra2025aes}. The resulting scores are presented in Table~\ref{tab:waveform_recon}. This indicates that our DCAE does not surpass the reconstruction quality upper bound of 1D VAE.

\subsubsection{Song Generation Metrics}

For evaluating complete song generation, we assessed models including ACE-Step, Hailuo, Suno v3, Udio v1, Yue, and DiffRhythm.\footnote{For DiffRhythm, we utilized version 1.2 (1m35s), which was its latest version updated on May 9, 2025.}

\paragraph{Style and Lyric Alignment.}
Style alignment measures how well generated music matches the specified style. It is quantified using CLAP and Mulan scores. For the CLAP score, we calculate the similarity between the text vector of the input tags and the audio vector using the implementation from stable-audio-metrics~\cite{stable-audio-metrics}. To complement this, and address potential limitations of using a non-music-specific CLAP model, we also employed Mulan~\cite{zhu2025muq}. Mulan is a similar contrastive language-audio pretraining model but features an audio encoder with enhanced music understanding capabilities, thus providing a more reliable reference for style alignment. As shown in Figure~\ref{fig:style_lyric_scores}(a), Udio v1 achieved the highest combined style alignment, followed by Hailuo and then ACE-Step. Suno v3, Yue, and DiffRhythm ranked subsequently.

Lyric alignment was assessed by the average confidence of Whisper forced alignment. Specifically, we first process the lyrics by removing structural tags and non-phonetic punctuation. The cleaned lyrics are then tokenized using Whisper's tokenizer. We then obtain the hidden states for these text tokens. By computing the cross-attention scores between these text hidden states and the audio's hidden states, a dynamic programming algorithm derives a confidence score for each force-aligned text token. A higher average confidence across tokens signifies that, from Whisper's perspective, the sung lyrics are clearer and more accurately aligned. The specific implementation for forced alignment was based on~\cite{stable-ts}. Figure~\ref{fig:style_lyric_scores}(b) indicates that Hailuo demonstrated the best lyric alignment. DiffRhythm and ACE-Step also performed well in this regard, while Udio v1, Suno v3, and Yue achieved lower scores. It is noted that this metric might favor genres with sparse instrumentation.

\begin{figure}[H]
  \centering
  \begin{subfigure}[b]{0.49\textwidth}
    \centering
    \includegraphics[width=\linewidth]{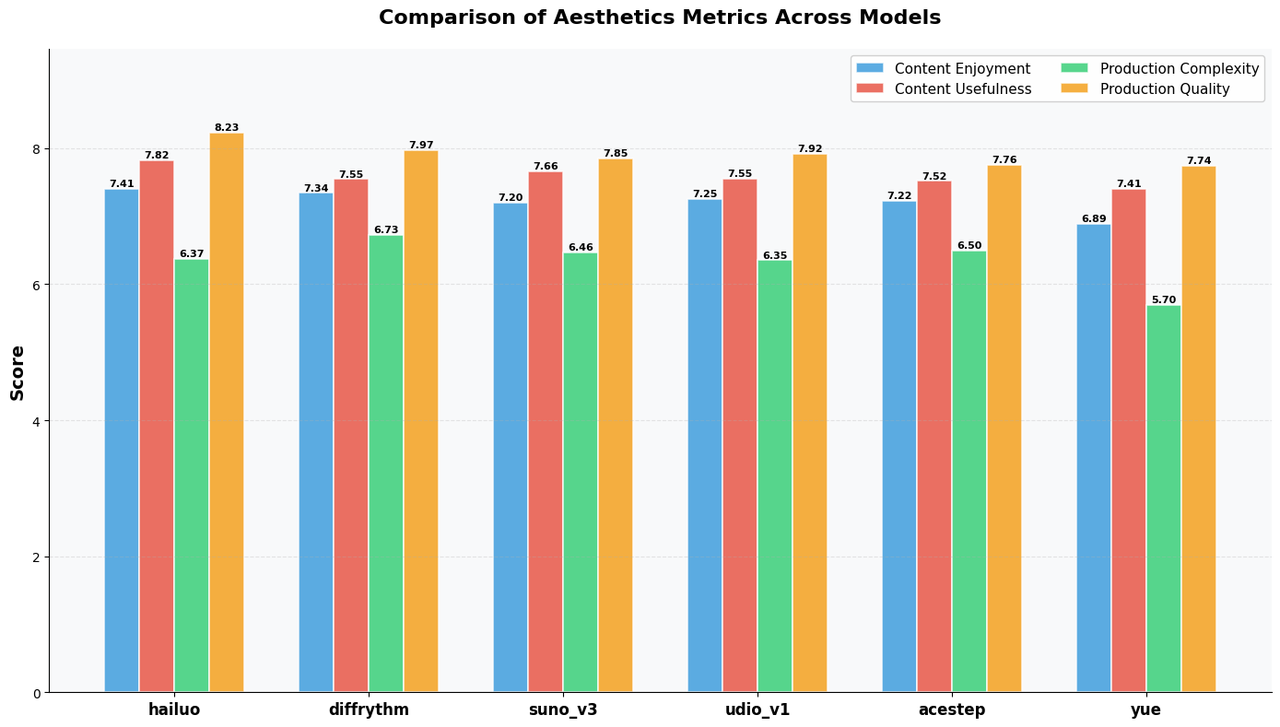} 
    \caption{Aesthetics Metrics}
    \label{fig:aesthetics_sub_metrics}
  \end{subfigure}
  \hfill 
  \begin{subfigure}[b]{0.49\textwidth}
    \centering
    \includegraphics[width=\linewidth]{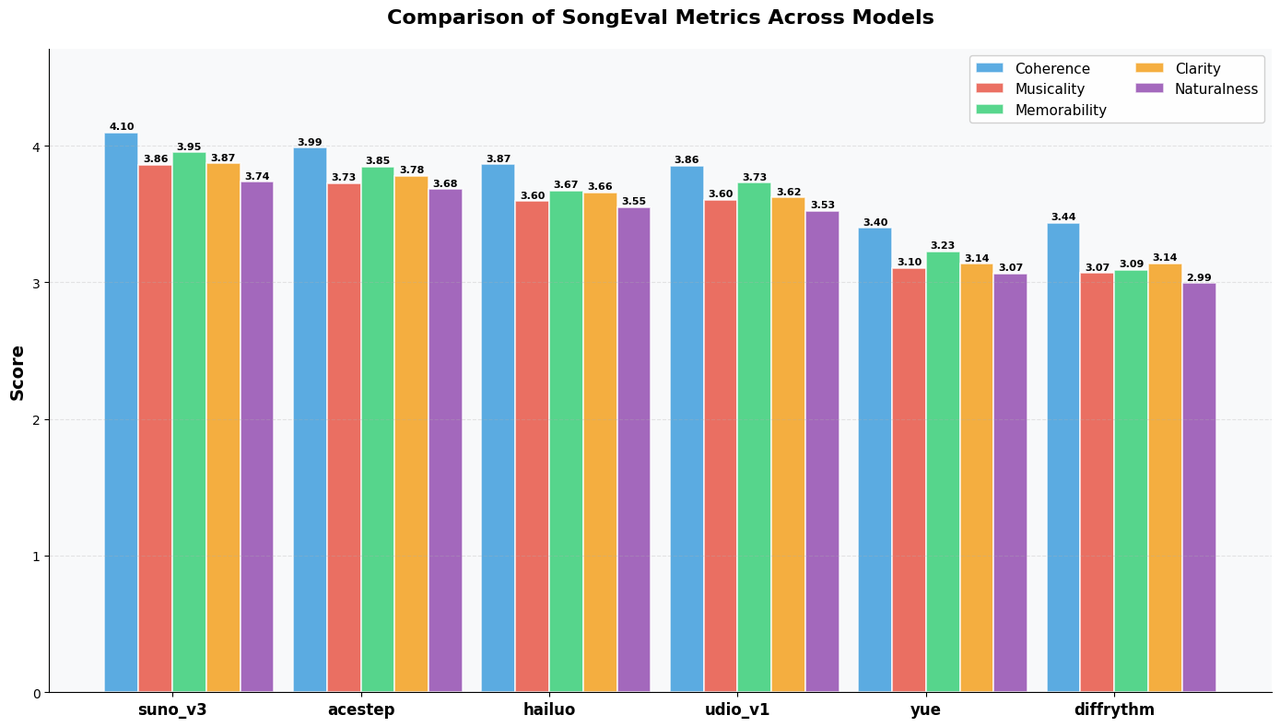} 
    \caption{SongEval Metrics.}
    \label{fig:songeval_sub_metrics}
  \end{subfigure}
  \caption{Comparison of (a) Audiobox Aesthetics Metrics and (b) SongEval Musicality Metrics Across Models.}
  \label{fig:quality_musicality_scores}
\end{figure}

\begin{table}[H]
\centering
\caption{Generation Speed Comparison on RTX 4090. RTF (Real-Time Factor) - higher is faster.}
\label{tab:speed_test}
\begin{tabular}{lrr}
\toprule
Method & RTF & Relative Speed \\
\midrule
Yue & 0.083x & 1.00x \\
DiffRhythm & 10.03x & 120.84x \\
ACE-Step & 15.63x & 188.31x \\
\bottomrule
\end{tabular}
\end{table}

\paragraph{Aesthetic Quality and Musicality.}
Aesthetic quality was evaluated using Meta's Audiobox framework across four dimensions. As detailed in Figure~\ref{fig:quality_musicality_scores}(a), Hailuo showed strong performance across all dimensions, with DiffRhythm also performing well. Suno v3 and Udio v1 yielded competitive scores, while ACE-Step positioned in the mid-to-upper range. Yue generally scored lowest. These scores can be style-dependent.

For musicality, we employed SongEval~\cite{yao2025songeval}, which is based on the first large-scale, open-source dataset for human-perceived song aesthetics. The SongEval toolkit enables automatic scoring of generated songs across five perceptual aesthetic dimensions designed to align with professional musician judgments. Extensive evaluations have demonstrated that SongEval's metrics exhibit high consistency with human subjective preferences, achieving over 90\% agreement at both utterance-level and system-level, significantly higher than the 60-70\% reported for AudioBox-aesthetics in similar comparisons. This indicates a greater reliability of SongEval in reflecting human perception of musicality. Assessed by SongEval across these five dimensions (see Figure~\ref{fig:quality_musicality_scores}(b)), Suno v3 achieved the highest Coherence and strong overall performance. ACE-Step demonstrated competitive results, particularly in Memorability and Clarity. Hailuo and Udio v1 also exhibited robust musicality, while Yue and DiffRhythm scored comparatively lower.

\paragraph{Generation Speed.}
Generation speed was benchmarked on consumer-grade hardware (NVIDIA RTX 4090 GPU) for Yue, DiffRhythm, and ACE-Step. Performance was measured using the Real-Time Factor (RTF), where a higher RTF indicates faster generation (e.g., RTF of 2x means twice as fast as real-time). As detailed in Table~\ref{tab:speed_test}, ACE-Step achieves the fastest generation with an RTF of 15.63x, followed by DiffRhythm at 10.03x. Yue is the slowest among the three methods with an RTF of 0.083x, meaning it generates audio at only 8.3\% of real-time speed. ACE-Step is approximately 188 times faster than Yue, while DiffRhythm is approximately 121 times faster than Yue.

\subsection{Metrics Interpretation}
Our evaluation reveals several critical observations regarding ACE-Step's performance in comparison to contemporary music generation systems. 

First, we identify a substantial divergence between subjective human evaluations and objective automatic metrics. This discrepancy arises from two primary factors: (1) the inherently subjective nature of musical perception, where individual preferences introduce variability in human assessments, and (2) persistent limitations in existing objective metrics to comprehensively capture perceptual qualities. While certain technical indicators demonstrate correlation with human judgment, others exhibit inconsistencies or fail to account for nuanced musical attributes. Notably, ACE-Step achieves state-of-the-art performance among open-source models on SongEval, the most reliable objective metric demonstrating closest alignment with human evaluations.

Second, comparative analysis with open-source alternatives reveals ACE-Step's superior capabilities across key musical dimensions. As shown in Figure~\ref{fig:genre_sub_score}, our model demonstrates significantly higher genre fidelity scores compared to competing approaches. Similarly, Figure~\ref{fig:songeval_sub_metrics} illustrates ACE-Step's advantages in overall musicality according to SongEval metrics. While Audiobox's aesthetic scores (Figure~\ref{fig:aesthetics_sub_metrics}) suggest comparable performance, these metrics appear less sensitive to perceptual quality differences in cross-model comparisons. Notably, DiffRhythm's exceptional lyric alignment scores (Figure~\ref{fig:lyric_sub_score}) likely stem from its pre-aligned lyric integration strategy, which bypasses cross-attention mechanisms by directly embedding lyrical content based on LLM-generated timestamps. However, this architectural choice correlates with compromised musical integration, as evidenced by DiffRhythm's comparatively lower SongEval scores - potentially resulting from rigid vocal placement that lacks contextual adaptability. Additionally, ACE-Step's multilingual support (19 languages) introduces training complexity trade-offs compared to monolingual systems, as fixed computational budgets necessitate distributed language-specific optimization.

Third, generation speed comparisons presented in Table~\ref{tab:speed_test} reveal unexpected efficiency advantages for ACE-Step (3.5B parameters) over the smaller DiffRhythm (1B parameters). This efficiency likely originates from linear transformer architecture benefits combined with implementation-specific optimizations. In contrast, Yue's generation speed proves impractical for interactive applications, exhibiting both excessive latency and strict input formatting requirements that limit usability.

Finally, we observe systematically lower performance metrics for DiffRhythm and Yue compared to their original publications. We attribute this difference to methodological variations: while these works employ audio prompts during evaluation, our protocol exclusively utilizes textual conditioning without audio references. We argue this approach establishes a more rigorous and realistic evaluation framework that better reflects typical usage scenarios requiring text-to-music generation without audio priors.

Collectively, these findings position ACE-Step as a leading open-source solution for multilingual music generation, while highlighting critical research directions for advancing both model architectures and evaluation methodologies in music AI research.
\section{Application}
\label{sec:application}

\begin{figure}[htbp]
    \centering
    \includegraphics[width=0.6\textwidth]{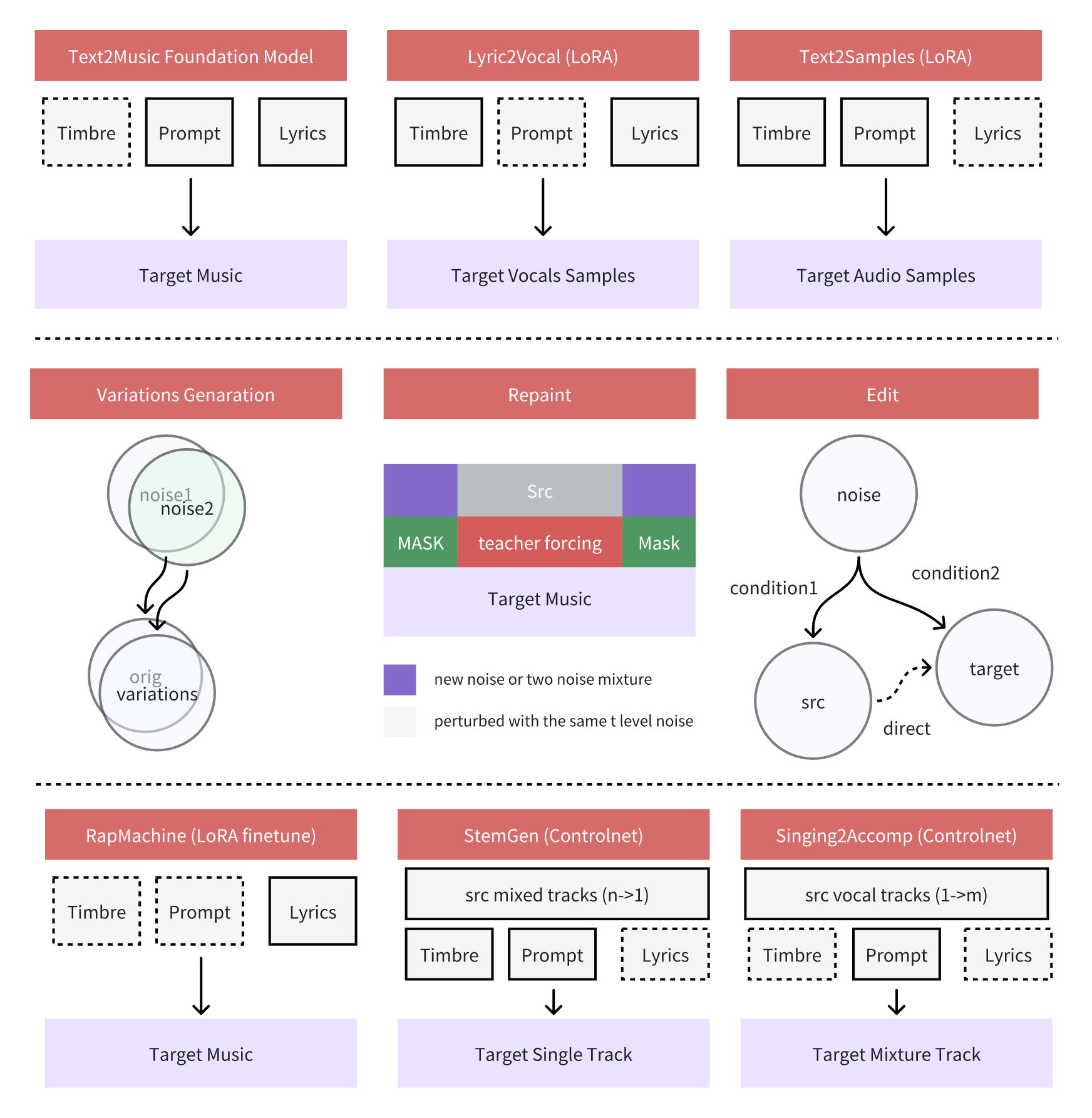} 
    \caption{An overview of the controllable features and specialized downstream applications of our model.}
    \label{fig:application_map}
\end{figure}

Our proposed model offers a suite of controllable features and downstream applications, demonstrating its versatility and practical utility in music generation and manipulation. These capabilities empower users with fine-grained control over the generation process and enable targeted modifications to existing audio, as well as specialized generation tasks through fine-tuning. An overview of these functionalities is presented in Figure~\ref{fig:application_map}.

\subsection{Core Controllability and Editing Features}
\label{ssec:controllability_editing}
The model incorporates several advanced techniques for controllable audio synthesis and direct editing of musical content.

\paragraph{Variations Generation} is supported through training-free, inference-time optimization. This process initiates with noise from a flow-matching model, to which additional Gaussian noise is introduced based on trigFlow's noise formulation ~\cite{lu2025simplifyingstabilizingscalingcontinuoustime}. The degree of variation is controlled by adjusting the mixing ratio between the original and new noise, allowing for a spectrum of modifications.

\paragraph{Audio Repainting} enables targeted modifications by introducing noise to specific regions of an audio input and applying mask constraints during the ODE solving process. This allows for altering specific aspects (e.g., instrumental lines) while preserving the rest of the audio. Repainting can also be combined with variations generation for localized changes in style, lyrics, or vocals.

\paragraph{Lyric Editing}~\cite{kulikov2024flowedit} leverages flow-edit technology for localized lyrical modifications while preserving melody, vocal characteristics, and accompaniment. This feature operates on both generated and uploaded audio, significantly enhancing creative possibilities. While currently optimized for small segment edits to prevent distortion, multiple sequential edits can achieve substantial lyrical revisions.

\subsection{Specialized Generation Tasks via Fine-tuning}
\label{ssec:specialized_generation}
To further demonstrate the model's adaptability, we explored fine-tuning using Low-Rank Adaptation (LoRA) and ControlNet-LoRA architectures for specialized generation tasks.

\paragraph{Lyric-to-Vocal} is achieved using a LoRA module fine-tuned on pure vocal recordings. This model directly synthesizes vocal samples from textual lyrics, proving useful for creating vocal demos, guide tracks, and aiding in songwriting and vocal arrangement experimentation by quickly auditioning lyrical ideas.

\paragraph{Text-to-Sample} employs a LoRA fine-tuned on pure instrumental and audio sample data. It can generate conceptual music production samples, such as instrument loops and sound effects, from descriptive text prompts, thereby streamlining the workflow for producers and sound designers.
\paragraph{StemGen} utilizes a ControlNet architecture \cite{zhang2023addingconditionalcontroltexttoimage} fine-tuned on multi-track audio data for the conditional generation of individual instrument stems. It accepts a reference audio track and a specification for the desired instrument (either as a textual label or a reference audio snippet) as input. The model then outputs an instrumental stem that is harmonically and rhythmically coherent with the reference track, such as generating a piano accompaniment for a given flute melody or synthesizing a jazz drum pattern for a lead guitar.

\paragraph{Rap Machine} is a LoRA module fine-tuned on a curated dataset of rap and hip-hop music, with a particular emphasis on Chinese content after rigorous data processing. This specialization significantly improves Chinese lyrical pronunciation, enhances adherence to hip-hop and electronic music conventions, and promotes diverse vocal performances. Beyond its primary focus, it enables the creation of varied hip-hop tracks and can be blended with other genres to introduce distinct vocal details or cultural flavors. Control over vocal timbre and performance techniques allows for further output refinement. This LoRA serves as a proof-of-concept, showcasing the foundation model's capability to generate novel styles and enhance linguistic performance via targeted fine-tuning.

\paragraph{Singing-to-Accompaniment} conceptually an inverse operation to StemGen, focuses on producing a complete instrumental accompaniment from an isolated vocal track, resulting in a mixed master track. It takes a monophonic vocal input and a user-specified musical style (e.g., "pop ballad," "acoustic folk") to generate a full instrumental backing. This capability allows for the creation of rich instrumental arrangements that musically support the input vocals, offering a streamlined approach for artists to produce professional-sounding accompaniments.
\section{Discussion}
\label{sec:discussion}

We introduced ACE-Step, an open-source foundation model for music generation. Our work successfully demonstrates the viability of our proposed architecture, which integrates a diffusion-based generator with Sana’s Deep Compression AutoEncoder (DCAE) and a lightweight linear transformer. This represents a significant step towards realizing a versatile music generation foundation model, akin to what Stable Diffusion~\cite{rombach2021highresolution} has achieved for image generation. We believe the current model serves as a robust proof-of-concept and a crucial stepping stone. Through continuous iteration and development, we are confident that the gap between open-source and closed-source models in music generation can be progressively narrowed.

Despite its achievements, ACE-Step has several limitations that we plan to address in future work:

\begin{itemize}
    \item \textbf{Upper Bound on Audio Quality:} The current audio quality is inherently limited due to two primary factors.
    \begin{itemize}
        \item Firstly, our approach relies on a mel-spectrogram-based DCAE and a 32kHz monophonic vocoder, rather than a direct end-to-end audio-to-audio generation pipeline. This choice, while efficient, can restrict the fidelity of the output.
        \item Secondly, during training, all audio data was uniformly resampled to 44.1kHz. A more optimal approach would involve using a detection script to filter and utilize genuinely high-resolution music for training. Alternatively, a multi-stage training paradigm could be adopted: initially training on lower-resolution data to learn structural aspects, followed by fine-tuning on high-resolution data to capture finer acoustic details.
    \end{itemize}

    \item \textbf{Style and Lyric Adherence:} While ACE-Step demonstrates commendable performance, there is room for improvement in its adherence to specified musical styles and lyrical content. Our initial implementation of style tagging, for instance, could be refined for better accuracy and granularity, leading to more precise stylistic control.

    \item \textbf{Variability in Generation Quality:} The generation process sometimes exhibits a "gacha-like" (lottery-style) experience, where obtaining optimal results may require multiple attempts. This indicates a degree of variability in output quality that we aim to reduce.
\end{itemize}

To address these limitations and further advance the capabilities of ACE-Step, we outline the following directions for future work:

\begin{itemize}
    \item \textbf{Data Augmentation and Extended Training:} We plan to significantly increase the volume of training data and extend the training duration to enhance the model's generalization capabilities and overall performance.

    \item \textbf{Improved Multilingual and Multi-Style Support:} Future iterations will focus on better support for diverse musical styles and languages, with an emphasis on improving the alignment between textual prompts (lyrics, style descriptions) and the generated audio.

    \item \textbf{Advanced 1D VAE Integration:} We intend to explore and integrate more sophisticated 1D Variational Autoencoders (VAEs) for audio representation. A better VAE could lead to improved compression and reconstruction quality, directly benefiting the final audio output.

    \item \textbf{Reinforcement Learning for Musicality:} We aim to incorporate reinforcement learning (RL) techniques to reward the model for generating music that exhibits higher musicality and aligns more closely with human preferences. This could involve training reward models based on human feedback.

    \item \textbf{Native Support for Mask-Aware Training:} The current "extend" functionality (for music continuation or inpainting) is not natively integrated into the training process. We plan to implement mask-aware training, which would allow the model to learn these tasks end-to-end. This is expected to significantly improve the quality of music continuation (outpainting) and inpainting (in-filling).
\end{itemize}

By pursuing these avenues, we aim to further refine ACE-Step, pushing the boundaries of open-source music generation and solidifying its role as a foundational tool for artists, producers, and researchers.
\newpage
\section*{Acknowledgements} 
\label{sec:acknowledgements}

This project is co-led by ACE Studio and StepFun.

We extend our profound gratitude to StepFun for their generous provision of computing power and storage infrastructure. Without such critical support, the development of ACE-Step would not have been possible. We are also deeply thankful to their team for organizing and conducting the subjective evaluations of our model.

We are immensely grateful to Jing Guo and Sean Zhao for their intensive involvement in decision-making and discussions concerning the training and application aspects of ACE-Step. Their meticulous weighing of trade-offs was instrumental in shaping ACE-Step into the versatile foundation model it has become.

Our sincere thanks go to Sen Wang and Shengyuan Xu for their invaluable support in data engineering. Effective data management was foundational to our efforts, and ACE-Step could not have been successfully trained without their expertise.

Finally, we wish to express our deepest appreciation to Junmin Gong for his extensive contributions, which encompassed algorithm development, framework design, conducting experiments, model training, and evaluation. Furthermore, we thank him for authoring the report, preparing the open-source code release, and creating the project webpage.
\newpage
\bibliographystyle{unsrt}
\bibliography{references}  

\newpage
\appendix
\section{Training Details}
\label{sec:appendix_training_details} 

\begin{table}[htbp]
    \centering
    \caption{Music DCAE Training Details.}
    \label{tab:dcae_training_details}
    \begin{tabularx}{\textwidth}{@{}lX@{}}
        \toprule
        Parameter           & Specification \\
        \midrule
        Dataset             & Entirety of our 100,000-hour curated music dataset. \\
        Latent Space        & Encoder produces a latent representation with a temporal dimension of 128 frames for input segments of approx. 11.88 seconds. \\
        Hardware            & 120 NVIDIA A100 GPUs. \\
        Training Steps      & 140,000 steps. \\
        Batch Size (Global) & 480 (4 per GPU). \\ 
        Total Duration      & Approx. 5 days. \\
        \addlinespace 
        Discriminators      & Multi-discriminator GAN setup for high-quality reconstruction:
                            \begin{itemize}[noitemsep, topsep=0pt, partopsep=0pt, leftmargin=*]
                                \item Standard patch-based discriminator with spectral normalization.
                                \item StyleGAN's Discriminator2DRes.
                                \item SwinDiscriminator2D.
                            \end{itemize} \\
        \addlinespace
        Training Stages     & \begin{itemize}[noitemsep, topsep=0pt, partopsep=0pt, leftmargin=*]
                                \item \textbf{Phase 1 (1 day):} Initial training focused solely on reconstruction (MSE loss).
                                \item \textbf{Phase 2 (4 days):} Encoder weights frozen; decoder trained with combined MSE and adversarial losses.
                            \end{itemize} \\
        \bottomrule
    \end{tabularx}
\end{table}

\begin{table}[htbp]
    \centering
    \caption{ACE-Step Model Training Details.}
    \label{tab:ace_step_training_details}
    \begin{tabularx}{\textwidth}{@{}lX@{}}
        \toprule
        Parameter                   & Specification \\
        \midrule
        Computational Resources     & 15 nodes, each with 8 NVIDIA A100 GPUs (120 GPUs total). \\
        Batch Size                  & Per-GPU: 1, Global: 120. \\
        Steps                 & 460,000 steps for pretrain; 240,000 steps for finetune. \\
        Dataset               & Full 100,000-hour music dataset for pretrain; Curated high-quality subset of approx. 20,000 hours for fintune\\
        Total Training Time     & Approx. 264 hours. \\
        Optimizer Type        & AdamW with 1e-2 weight decay and (0.8, 0.9) betas. \\
        Gradient Clipping     & Max norm of 0.5. \\
        Learning Rate         & 1e-4, with linear warm-up over the first 4,000 steps. \\
        Timestep Sampling     & Logit-normal scheme: $u \sim \mathcal{N}(\text{mean}=0.0, \text{std}=1.0)$; shift=3.0. \\
        Sequence Lengths (Max Context) & 4096 tokens for lyrics, 256 tokens for text prompts, 2584 tokens for mel-spectropgram latents\\
        Conditional Dropout         & \begin{itemize}[noitemsep, topsep=0pt, partopsep=0pt, leftmargin=*]
                                        \item Global conditioning: 15\% dropout rate.
                                        \item Lyrics \& Prompt embeddings: 15\% dropout rate each (independent).
                                        \item Speaker embedding: 50\% dropout rate (independent).
                                    \end{itemize} \\
        \bottomrule
    \end{tabularx}
\end{table}

\newpage
\begin{CJK}{UTF8}{gbsn}

\section{Evaluation Prompts}
\label{sec:eval_prompts_appendix}

The following 20 prompts were used for the song generation tasks described in Section 5.2. These prompts cover diverse musical styles and include both English and Chinese lyrics, generated by an LLM. Each prompt consists of style tags and lyrics.

\subsection*{Prompt 1 (test001.txt)}
\begin{verbatim}
## Prompt
hiphop, rap, trap, boom bap, old school, gangsta rap, east coast, west
coast

## Lyrics
[verse]
街头的风拂过深夜的心
脚步嗒嗒像打鼓的音
霓虹灯下闪烁我的身影
城市喧嚣里找自由的魂灵


[verse]
旧城墙上乱画的图腾
解析梦想燃烧的过程
脚下的路像谜题的分层
解不开却又不敢不狂奔


[chorus]
唱出我的人生旋律
每个音符都是真实的轨迹
前路再崎岖 腰板也要挺直
就算跌倒也要笑笑继续


[verse]
地铁车厢里人潮的涌动
耳机里的B-box是我的放纵
用音符写下每一场感动
让故事流淌像心绪般翻涌


[bridge]
现实像风把我吹得凌厉
但内心深处还有火种延续
用歌词点燃每一个深夜
把希望种在破碎的心灵


[chorus]
唱出我的人生旋律
每个音符都是真实的轨迹
前路再崎岖 腰板也要挺直
就算跌倒也要笑笑继续

\end{verbatim}

\subsection*{Prompt 2 (test002.txt)}
\begin{verbatim}
## Prompt
pop, dance pop, synthpop, bubblegum pop, k-pop, j-pop, electropop

## Lyrics
[verse]
夜晚的星空在诉说
藏在心底的梦和火
想要追逐不愿停泊
让光芒照亮沉默的我


[verse]
流星划过像个谜题
带着秘密闯入梦里
许下心愿毫不犹豫
哪怕前路荆棘密密


[chorus]
让我飞越云层去寻找
用双手拥抱世界的微笑
不管天多高梦多渺小
我会让希望在黑暗燃烧


[bridge]
回忆像风轻轻掠过
吹动心跳不能退缩
未来的路就算寂寞
也要高歌不停放纵洒脱

[verse]
星星闪烁似在低语
命运从不轻易放弃
越是挣扎越是美丽
就算跌倒也会铭记

[chorus]
让我飞越云层去寻找
用双手拥抱世界的微笑
不管天多高梦多渺小
我会让光辉在黑暗燃烧

\end{verbatim}

\subsection*{Prompt 3 (test003.txt)}
\begin{verbatim}
## Prompt
rock, classic rock, hard rock, alternative rock, indie rock, punk rock,
garage rock

## Lyrics
[verse]
天空压下来
沉重的黑
风声呼啸
像是愤怒的泪
谁在挣扎
谁在咆哮
看不清的脸
在深夜燃烧


[verse]
脚下的路
崩裂成灰
心中的火
灼烧着肺
想要逃离
却被困住
嘶吼的声音
没人能藏住


[chorus]
黑夜的呐喊
不会停息
撕裂这世界
唤醒自己
孤独的灵魂
在废墟游荡
不再沉默
不再隐藏


[verse]
影子追逐着
破碎的梦
眼神迷离
仿佛无尽痛
欲望吞噬
破坏重生
在深渊中重建一座城


[bridge]
挣脱枷锁
放下虚伪
烈火焚尽
那些虚妄的美
心跳加速
电流流窜
点燃这一切
直到那尽头


[chorus]
黑夜的呐喊
不会停息
撕裂这世界
唤醒自己
孤独的灵魂
在废墟游荡
不再沉默
不再隐藏

\end{verbatim}

\subsection*{Prompt 4 (test004.txt)}
\begin{verbatim}
## Prompt
folk, acoustic folk, indie folk, traditional folk, americana, singer-
songwriter

## Lyrics
[verse]
山间小路弯又长
晨雾低垂掩山岗
脚步轻踏叶儿响
风吹草木诉过往


[verse]
鸟儿啼声穿林去
溪水叮咚流不停
微光透过松树影
映出岁月的倒影


[chorus]
山高水长路悠远
心中故乡梦绵长
背起行囊走四方
歌声伴我过流年


[verse]
村头炊烟轻轻扬
小孩追逐笑声长
老树根旁旧时光
回忆满怀酒一觞


[verse]
星光点点夜未央
月亮挂在那树上
倦鸟归巢影成双
故土情深伴梦乡


[chorus]
山高水长路悠远
心中故乡梦绵长
背起行囊走四方
歌声伴我过流年

\end{verbatim}

\subsection*{Prompt 5 (test005.txt)}
\begin{verbatim}
## Prompt
rnb, rhythm and blues, soul, neo soul, contemporary rnb, funk

## Lyrics
[verse]
灯火在街角晃动的影子
心跳和人群节奏很一致
车水马龙像流动的诗句
这一刻我只想见到你


[verse]
钢铁森林藏着许多秘密
风吹过耳边带着回忆
我的脚步追逐你的气息
整座城市却默契藏起你踪迹


[chorus]
我的城市太大装不下孤单
人潮汹涌却像空荡的展览
我想问问你会不会也感叹
在这迷宫里找不到答案


[bridge]
路灯的光亮像星星闪烁
它在指引却又迷惑
是不是我们注定错过
还是会在转角遇见你我


[verse]
时间在流逝偷走了勇气
手心的温度慢慢冷去
人群中每个背影的痕迹
却偏偏不是你熟悉的气息


[chorus]
我的城市太大装不下孤单
人潮汹涌却像空荡的展览
我想问问你会不会也感叹
在这迷宫里找不到答案

\end{verbatim}

\subsection*{Prompt 6 (test006.txt)}
\begin{verbatim}
## Prompt
punk, punk rock, hardcore punk, pop punk, post-punk, skate punk

## Lyrics
[verse]
月光像刀划破夜的天
城市在嘶吼没人入眠
黑白的规则我全都厌倦
我要自由不受限


[chorus]
燃烧吧青春像火焰在飞
别让世界把你心再锁一回
我们是流星划破的轨
燃尽所有偏不后退


[verse]
墙上的涂鸦没人去擦
脚下的街道在自由发芽
每个怒吼都藏着希望的花
让平庸被分崩瓦解


[chorus]
燃烧吧青春像火焰在飞
别让世界把你心再锁一回
我们是流星划破的轨
燃尽所有偏不后退


[bridge]
听那节奏像心跳在吵
嘶哑的呐喊击碎沉默的牢
这一刻世界只属于咆哮
规则在燃烧碎片在飘


[chorus]
燃烧吧青春像火焰在飞
别让世界把你心再锁一回
我们是流星划破的轨
燃尽所有偏不后退
\end{verbatim}

\subsection*{Prompt 7 (test007.txt)}
\begin{verbatim}
## Prompt
electronic, edm, house, techno, trance, synthwave, ambient

## Lyrics
[verse]
霓虹灯光撕扯这夜空
城市脉搏跳动像梦中
每一步都像心跳的节拍
灵魂跟着节律不再徘徊


[verse]
街道在闪耀像银河的碎片
人群如波浪翻涌着瞬间
手指轻触那温暖的光线
所有烦恼被丢在了昨天


[chorus]
这是闪耀的夜晚别停下来
让音乐带着我们去腾空
跳跃在这梦幻的舞场
让音符把孤单全埋葬


[verse]
眼前世界像一场旋转梦
星光洒满心底的每个空
无数可能在指尖中放纵
在这夜里未来都被填充


[bridge]
听 那声响刺破了时间
看 那光芒点亮了明天
一切都在这一刻交叠
自由的心灵没有界限


[chorus]
这是闪耀的夜晚别停下来
让音乐带着我们去腾空
跳跃在这梦幻的舞场
让音符把孤单全埋葬

\end{verbatim}

\subsection*{Prompt 8 (test008.txt)}
\begin{verbatim}
## Prompt
jazz, smooth jazz, bebop, cool jazz, fusion, latin jazz

## Lyrics
[verse]
月光洒在窗前
月影婆娑
风儿低语细声
似诉衷情
街灯摇曳朦胧
映出你轮廓
心中泛起涟漪
如梦似幻真情


[verse]
你的笑容如星
闪烁在夜空
听那轻声细语
像一首老歌
时间仿佛停驻
片刻也无痕
这一刻的温存
让人心醉沉沦


[chorus]
哦
夜晚的呢喃
诉说着爱恋
在每个音符间
轻拨心弦
你的眼眸如诗
写满了缠绵
让我在这旋律里
与你共缠绵


[bridge]
夜色深沉如墨
思绪随风飘
你的影子渐远
心却停不了
在这温柔世界
所有都轻描
只愿与你相伴
把时光忘掉


[verse]
每一声叹息
都藏着期盼
像微风掠过树叶
轻轻作伴
你的名字回荡
夜空多灿烂
在这静谧时分
爱意如浪漫


[chorus]
哦
夜晚的呢喃
诉说着爱恋
在每个音符间
轻拨心弦
你的眼眸如诗
写满了缠绵
让我在这旋律里
与你共缠绵
\end{verbatim}

\subsection*{Prompt 9 (test009.txt)}
\begin{verbatim}
## Prompt
reggae, roots reggae, dancehall, dub, ska, reggae fusion

## Lyrics
[verse]
清风掠过发梢的温柔
白沙铺向海天的尽头
棕榈叶跳起华尔兹舞
节拍应和心跳的频率

[chorus]
跃入粼粼波光的怀里
自由是浪花轻吻脚底
咸涩的风哼蓝色小调
此刻星辰坠入你眼底

[verse]
七色桥横跨碧海银堤
热带果香漫过潮汐
赤足踏过滚烫的砂砾
脉搏随烈日沸腾跃起

[chorus]
跃入粼粼波光的怀里
自由是浪花轻吻脚底
咸涩的风哼蓝色小调
此刻星辰坠入你眼底


[bridge]
将烦恼藏进贝壳缝隙
脚印踩着爵士鼓点游移
晚霞泼洒油画般绮丽
灵魂绽放出翡翠枝桠


[chorus]
跃入粼粼波光的怀里
自由是浪花轻吻脚底
咸涩的风哼蓝色小调
此刻星辰坠入你眼底
\end{verbatim}

\subsection*{Prompt 10 (test010.txt)}
\begin{verbatim}
## Prompt
dj, turntablism, club mix, remix, electronic dance, beat juggling

## Lyrics
[verse]
地铁播报惊碎了街影
穿铆钉靴的姑娘踏破寂静
霓虹刺破八月潮湿的雾气
耳机线牵动心跳的涟漪


[verse 2]
地下通道萨克斯在呜咽
追光灯扫过潮湿的砖墙
石膏像突然眨动眼睛
铁皮箱震落二十年的锈迹


[chorus]
跳起来跳起来别停歇
这一刻只属于狂野
放开手放开脚一起飞
夜色中舞动最无畏


[bridge]
旋转的裙摆盛放成火焰
贝斯撕开黎明的裂缝
流浪猫跃上配电箱指挥
昨夜雨珠在电缆上颤栗


[verse]
人潮在暗河里沸腾漫溢
陌生人撞出银河的剖面
汗珠折射七种语言的叹息
警戒线缠住天鹅的脖颈


[chorus]
跳起来跳起来别停歇
这一刻只属于狂野
放开手放开脚一起飞
夜色中舞动最无畏
\end{verbatim}

\subsection*{Prompt 11 (test011.txt)}
\begin{verbatim}
## Prompt
blues, delta blues, chicago blues, electric blues, acoustic blues, blues
rock, soul blues, texas blues, boogie-woogie, gospel blues, british blues,
jump blues, harmonica blues

## Lyrics
[verse]
Train howls like my baby's gone
Steel tracks drown in her perfume
Moon stains rails with whiskey tears
Her ghost haunts the gathering gloom


[chorus]
Wheels scream, heart cracks slow
Laugh echoes in engine's glow
Gone, gone
Shadow clings to me
Midnight steel stole my baby


[verse]
Smoke climbs like my pain
Whistle shrieks her name
Miles bleed, blades in sound
Took love where tracks pound


[bridge]
Steel sings lonesome tune
Cold jealous moon
Scarred land's cruel brand
Ghost slips through my hand


[chorus]
Wheels scream, heart cracks slow
Laugh echoes in engine's glow
Gone, gone
Shadow clings to me
Midnight steel stole my baby


[verse]
Wild winds whip the tracks
Earth shakes like it knows
Train howls through the black
Leaves me 'neath stars' cracks
\end{verbatim}

\subsection*{Prompt 12 (test012.txt)}
\begin{verbatim}
## Prompt
classical, baroque, classical period, romantic era, symphony, chamber
music, opera, concerto, neoclassical, choral

## Lyrics
[chorus]
The river of time
It pulls us fast
No moment stays
They never last
Through fleeting days
Through moonlit skies
We chase the tide
We say goodbyes

[verse]
The stones below
They mark the way
Of stories told
Of yesterday
The water whispers
Soft and low
A melody only the river knows

[chorus]
The river of time
It pulls us fast
No moment stays
They never last
Through fleeting days
Through moonlit skies
We chase the tide
We say goodbyes

[bridge]
The current laughs
It doesn’t pause
It knows no fear
It breaks no laws
It carves the earth
It steals the years
It leaves us with our joys and tears

[chorus]
The river of time
It pulls us fast
No moment stays
They never last
Through fleeting days
Through moonlit skies
We chase the tide
We say goodbyes
\end{verbatim}

\subsection*{Prompt 13 (test013.txt)}
\begin{verbatim}
## Prompt
country, acoustic-guitar, storytelling, rural, heartland, twang, banjo,
fiddle, dobro, honky-tonk, western, cowboy, ballad, bluegrass, country-
rock, traditional-country, modern-country

## Lyrics
[verse]
Road's long but I'm not alone
Sky above I'm flesh and bone
Miles sing memories
Calling me home


[verse]
Through hills and deep valleys
Past willows' whispers
Wind carries tales
Back to the dust


[chorus]
Back to dust where wild grass grows
Earth speaks soft and river flows
Stars are my map moon my guide
Back to dust where heart won't hide


[verse]
Wore the city like a coat
Threads came loose seams just broke
Freedom here I can't deny
In the soil and open sky


[bridge]
Dirt on my hands feels like a hug
Each grain a story each mark a trace
Of a life that's raw a heart that's true
Back to the dust where I start anew


[chorus]
Back to dust where wild grass grows
Earth speaks soft and river flows
Stars are my map moon my guide
Back to dust where heart won't hide
\end{verbatim}

\subsection*{Prompt 14 (test014.txt)}
\begin{verbatim}
## Prompt
disco, transitional disco, electronic, rhythm, beat, bass, groove, funk,
70s, club, nightlife, party, upbeat, energetic, repetitive, catchy, synth,
drum-machine

## Lyrics
[verse]
Streetlights flicker on the rooftops
The jukebox coughs up vinyl aches
Patent shoes crack linoleum scars
Mirrorball sheds '79 silver


[chorus]
Moonlight bleeds through spinning gears
Hips unlock rusted frontier
Leather soles spark on concrete veins
We sweat this city alive again


[verse]
Polyester blooms salt constellations
Stranger's cologne burns Adam's apple
Smoke traces tango on sticky tiles
AC wheezes disco defibrillator


[bridge]
Rhine stones strand on collarbones
Talcum powder orbits slow
Needle skips a memory gun
Resurrects 3AM's neon sun


[chorus]
Moonlight bleeds through spinning gears
Hips unlock rusted frontier
Leather soles spark on concrete veins
We sweat this city alive again
\end{verbatim}

\subsection*{Prompt 15 (test015.txt)}
\begin{verbatim}
## Prompt
hiphop, rap, trap, boom bap, old school, gangsta rap, east coast, west
coast

## Lyrics
[verse]
Truth cuts clean
Steps carve unseen
Ash fuels rise
Eyes hold skies
[Chorus]
Raw truth flows
Soul volcanic
Tides meet wick
Unbreakable


[verse]
Severed puppet strings
Shallow tides recede
Phoenix bones take wing
Atmosphere I breathe


[bridge]
Shadowless stride
Dawn claws night
Veritas blade
Incandescent


[verse]
Doused flames burn brighter
Steel forged in thunder
Ventricles pump fire
Lightning sans plunder


[chorus]
Raw truth flows
Soul volcanic
Tides meet wick
Unbreakable
\end{verbatim}

\subsection*{Prompt 16 (test016.txt)}
\begin{verbatim}
## Prompt
jazz, smooth jazz, bebop, cool jazz, fusion, latin jazz

## Lyrics
[verse]
The moon's silver smile gleams
Whippoorwills sing in the twilight
Velvet whispers your name
Phantoms wear your scent


[chorus]
Moon, show the nightingales how to sing
June's passion burns bright but fades fast
Shadows gnaw at the dawn's leash
Heartbeats scatter like autumn leaves


[verse]
Stars flicker like Morse code
The saxophone exhales a mist of smoke
Your silhouette melts away
In the midnight groove of vinyl


[bridge]
Midnight's needle spins a tale
Dreams unravel into threads
Lovers' maps fade to moth-wing parchment
Memories drift like fallen leaves


[chorus]
Moon, teach the nightingales to sing
June's fever burns bright but fades fast
Shadows gnaw at the dawn's leash
Heartbeats scatter like autumn leaves


[verse]
The clock's face bleeds mercury
Night slips off its velvet glove
Moths carve hieroglyphs of light
The moon's grin sharpens
\end{verbatim}

\subsection*{Prompt 17 (test017.txt)}
\begin{verbatim}
## Prompt
metal, heavy-metal, electric-guitar, distortion, power-chords, drums,
double-bass, high-energy, aggressive, theatrical, anthemic, headbanging

## Lyrics
[verse]
Dark skies above
Endless torment
Shattered worlds
Souls are dormant
Blood-soaked ground
Cries of despair
The abyss calls
None are spared

[verse]
Chains of fire
Bind the weak
Silent screams
No words to speak
Chaos reigns
Death takes hold
Life extinguished
Hearts grow cold

[chorus]
Rise from ash
Feed the flame
Drown in darkness
Curse your name
Torn asunder
Shattered pride
In the abyss
No one hides

[verse]
Winds of sorrow
Endless pain
Drenched in blood
A crimson rain
No escape
No way to flee
Bound forever
Misery

[bridge]
Rage consumes
No light remains
Shattered bones
Eternal chains
Break the silence
Let it roar
Unleash the beast
Crave for more

[chorus]
Rise from ash
Feed the flame
Drown in darkness
Curse your name
Torn asunder
Shattered pride
In the abyss
No one hides
\end{verbatim}

\subsection*{Prompt 18 (test018.txt)}
\begin{verbatim}
## Prompt
pop, dance pop, synthpop, bubblegum pop, k-pop, j-pop, electropop

## Lyrics
[verse]
I found my shoes by the riverside
They lay there, silent, like forgotten dreams
The city sleeps but I stay awake
Chasing dreams with every step I take

[verse]
Neon signs hum a lullaby
But I’m flying high
No need to try
The ground is lava
I’m on a roll
Dancing on the moonlight
I’ve lost control

[chorus]
Oh oh
Feel the night in my veins
Oh oh
Breaking free from the chains
Oh oh
Let’s rewrite the tune
We’re dancing on the moon

[verse]
The air is thick with electric heat
I’m skipping beats with my reckless feet
The echoes sing in a rhythm raw
A song of chaos breaking every law

[bridge]
We’ve got no rules
We’ve got no time
The world spins fast but we climb
With every spark
With every sway
We steal the stars and light our way

[chorus]
Oh oh
Feel the night in my veins
Oh oh
Breaking free from the chains
Oh oh
Let’s rewrite the tune
We’re dancing on the moon
\end{verbatim}

\subsection*{Prompt 19 (test019.txt)}
\begin{verbatim}
## Prompt
reggae, roots reggae, dancehall, dub, ska, reggae fusion

## Lyrics
[verse]
The ocean whispers secrets to the shore
Sun-kissed vibes
Who could ask for more
Feet in the sand
Heart starts to sway
Feel the rhythm carry us away

[verse]
Palm trees dance beneath the golden light
Stars above glow
Guiding through the night
Life’s a wave
Let it take us high
Catch the breeze
No need to question why

[chorus]
We’re chasing the echoes
The island calls
Breaking the silence
As the rhythm falls
Every heartbeat sings a brand-new tune
Under the sun and the silver moon

[bridge]
Feel the warmth as it kisses your face
Every moment here is a sweet embrace
Time slows down
Worries fade from view
It’s a paradise made for me and you

[verse]
Hear the laughter
Voices on the wind
Every soul here feels like kin
Colors burst
Painting skies of gold
In this magic
Let your heart be bold

[chorus]
We’re chasing the echoes
The island calls
Breaking the silence
As the rhythm falls
Every heartbeat sings a brand-new tune
Under the sun and the silver moon
\end{verbatim}

\subsection*{Prompt 20 (test020.txt)}
\begin{verbatim}
## Prompt
rock, classic rock, hard rock, alternative rock, indie rock, punk rock,
garage rock

## Lyrics
[verse]
Woke up screaming in a shadowed haze
Broken truth tearing at the edges of my days
Crawling through the wreckage of another lie
Hands clenched fists raised asking the sky why

[verse]
City lights flicker like a dying flame
Every face I see starts to look the same
Echoes of a promise shattered in the air
The weight of nothingness pulling me nowhere

[chorus]
Shattered glass dreams cut through my skin
Feel the fire burning deep within
Rise from the wreckage crawl through the ash
I'm not staying silent while my soul turns to trash

[verse]
Every heartbeat pounding like a battle drum
Fighting for a future that will never come
Tearing at the chains that try to hold me down
Searching for salvation in this lost and broken town

[bridge]
Oh the fire's coming it's melting the chains
Can't drown the fury or silence the pain
The harder they push the louder I scream
Won't let them steal my shattered glass dream

[chorus]
Shattered glass dreams cut through my skin
Feel the fire burning deep within
Rise from the wreckage crawl through the ash
I'm not staying silent while my soul turns to trash
\end{verbatim}

\end{CJK}
\end{document}